\documentclass[reprint,amsmath,amssymb,aps]{revtex4-2}
\usepackage{graphicx,bm}

\begin{document}
\title{Arbitrarily Precise Quantum Alchemy}

\author{Guido Falk von Rudorff}
 \email{guido.vonrudorff@univie.ac.at}
\affiliation{Faculty of Physics, University of Vienna, Kolingasse 14-16, 1090 Vienna, Austria}

\begin{abstract}
Doping compounds can be considered a perturbation to the nuclear charges in a molecular Hamiltonian. Expansions of this perturbation in a Taylor series, i.e. quantum alchemy, has been used in literature to assess millions of derivative compounds at once rather than enumerating them in costly quantum chemistry calculations. So far, it was unclear whether this series even converges for small molecules, whether it can be used for geometry relaxation and how strong this perturbation may be to still obtain convergent numbers. This work provides numerical evidence that this expansion converges and recovers the self-consistent energy of Hartree-Fock calculations. The convergence radius of this expansion is quantified for dimer examples and systematically evaluated for different basis sets, allowing for estimates of the chemical space that can be covered by perturbing one reference calculation alone. Besides electronic energy, convergence is shown for density matrix elements, molecular orbital energies, and density profiles, even for large changes in electronic structure, e.g. transforming He$_3$ into H$_6$. Subsequently, mixed alchemical and spatial derivatives are used to relax H$_2$ from the electronic structure of He alone, highlighting a path to spatially relaxed quantum alchemy. Finally, the underlying code (APHF) which allows for arbitrary precision evaluation of restricted Hartree-Fock energies and arbitrary-order derivatives is made available to support future method development.
\end{abstract}

\maketitle

\section{Introduction}
At the core of most quantum chemistry methods is the search for sufficiently good approximations of the total energy $E$ of a system, given the positions $\mathbf{R}_I$ and nuclear charges $Z_I$ of all $N$ nuclei, the net charge $Q$ and the spin $\sigma$. While the Schrödinger equation as foundation of such approximations is commonly applied to integer nuclear charges only, the equation itself is perfectly valid for any non-integer nuclear charge. Since only integer nuclear charges can be realised in experiments, the individual compounds have been mostly thought of as isolated points in chemical space. In the realm of quantum alchemy however, non-integer nuclear charges are considered as a conceptual tool to interpolate between compounds, as e.g. done by Wilson in 1962\cite{Wilson1962}. This allows to employ well-established methods such as perturbation theory in order to describe similar compounds if the electronic structure of one compound is already available. In the context of materials design or high-throughput-screening, such a tool is most valuable, as expensive recalculations of slightly different compounds can be avoided, often times with limited loss of accuracy only\cite{von_Rudorff_2020, Rudorff2020}.

To that end, several formulations have been developed, all based on the derivatives of energies or electron densities $\rho$ with respect to either the nuclear charges or the external potential: a Taylor expansion of the energy\cite{Lilienfeld2009}, derivatives from conceptual DFT\cite{Parr1995}, and Alchemical Perturbation Density Functional Theory (APDFT) which is based on density derivatives\cite{von_Rudorff_2020}. While some formulations even allow for a change in numbers of electrons\cite{Cardenas2011,Balawender2019}, this work considers isoelectronic changes only.

Even though these pen-and-paper formulations already allow for fundamental insights such as quasi-symmetries\cite{von_Rudorff_2021} in chemical space or tests of the near-sightedness of the electron density\cite{Fias2017}, the underlying assumption---the convergence of the Taylor series---has only been postulated and motivated. Numerical tests so far are limited by the numerical accuracy and the finite number of expansion terms that could be calculated. 
A prerequisite for widespread application of quantum alchemy methods is a reliable computation of the corresponding derivatives to demonstrate convergence.

For crystals, a perturbative approach similar in spirit is Density functional perturbation theory\cite{Baroni1987,Gonze1995} (DFPT). DFPT---although only applicable to DFT---has been mostly applied to lattice dynamics for DFT calculations\cite{Baroni2001}. Alchemical changes have been\cite{Marzari1994,Gironcoli1991} explored in DFPT with responses up to second order using pseudopotentials, but are ultimately limited by the need for a manual derivation of an implementable derivative expression for higher orders. 

These different approaches to quantum alchemy have been used for catalyst discovery\cite{Griego2020,Griego2018,Saravanan2017}, covalent bond energies\cite{von_Rudorff_2020, Balawender2018}, non-covalent interactions\cite{von_Rudorff_2020}, mixtures\cite{Marzari1994,Gironcoli1991,Alfe2000}, deprotonation energies\cite{Rudorff2020,Munoz2020}, adsorption energies\cite{AlHamdani2017}, and reactions\cite{Sheppard2010}.

In the context of classical molecular dynamics calculations, classical alchemy is used in thermodynamic integration schemes which use the conversion of one molecule into another to obtain free energy differences\cite{Straatsma1992}. In this case, the force field parameters are varied to bring the system smoothly from one potential energy surface (PES) to another. This standard tool in classical calculations so far has seen only little attention in ab initio applications with the exception of hybrid schemes\cite{Cheng_2009,Sulpizi2010,Cheng2014}, commonly applied to systems where force fields are known to fail\cite{Gittus2018,Liu2013a}. Contributing factors to this limited use so far are limited support by quantum chemistry codes for the calculation of the required derivatives (e.g. $\partial_Z E$) and little information about the convergence radius of a perturbative expansion in chemical space. These two factors constitute a vicious circle deterring research efforts from this field of substantial promise. This work aims at breaking this self-stabilising situation by providing tools and numerical evidence of the convergence of such expansions in chemical space, which hopefully sparks more research into computationally efficient approaches and suitable basis sets.

So far, alchemical derivatives such as $\partial_ZE$ or $\partial_Z \rho$ have been obtained either by finite differences\cite{Lilienfeld2009}, with coupled-perturbed approaches\cite{Geerlings2014}, Fukui functions\cite{Geerlings2014} or by hand-crafted analytical expressions\cite{Lesiuk2012} for lower orders. Analytical expressions are much more numerically stable, but need to be derived from scratch for any level of theory. (Central) finite differences evaluate the (costly) PES multiple times for small displacements and approximate the derivatives from a weighted sum of these calculations.
If the displacement $h$ is too small, the energy between such displacements varies too little and the finite precision of the energy evaluation is amplified by the $h^{-n}$ term already in the first few orders. If the displacement is too large, the derivative becomes inaccurate as the expression from finite differences is only accurate in the limit of small $h$. This limited precision prohibits an evaluation of the convergence of the expansion, as researchers are practically limited to the first 5-6 orders\cite{von_Rudorff_2020,Domenichini2020}.

More recently, another approach to obtaining derivatives of arbitrary programs has emerged, driven by machine learning efforts: automatic differentiation (AD). In this approach, a program like a quantum chemistry code is treated as a complex and deeply nested function where the derivative is evaluated using repeated application of the chain rule. Early work on applying AD to quantum chemistry codes\cite{TamayoMendoza2018} has been followed up upon more recently\cite{Abbott2021,Pavosevic2020} for more advanced methods. These referenced works explain in detail the substantial challenges to transform a code into a form that can be differentiated using AD libraries. However, even AD is limited by machine precision which still does not allow to go to high orders of perturbations. This work combines finite differences and arbitrary precision math to obtain highly accurate energy derivatives for use in quantum alchemy. Having highly accurate derivatives is key to explore the convergence behaviour and the convergence radius limiting quantum alchemy.

\section{Methods}
\subsection{Quantum alchemy}
The goal of quantum alchemy is to model the alchemical change from a reference molecule to a target molecule, where the two molecular Hamiltonians only differ in the nuclear charges. This is conveniently done by linear interpolation\cite{Lilienfeld2009,Kirkwood1935} of the corresponding Hamiltonians $\hat H_\textrm{r}$ and $\hat H_\textrm{t}$ for reference and target molecule, respectively
\begin{align}
    \hat H(\lambda) \equiv \lambda\hat H_\textrm{t}+ (1-\lambda)\hat H_\textrm{t}
\end{align}
where $\lambda$ parametrizes the path. Now the energy of the target molecule $E_\textrm{t}$ can be obtained from a Taylor expansion in $\lambda$.
\begin{align}
    E_\textrm{t} = \sum_{n=0}^\infty \frac{1}{n!}\frac{\partial^n}{\partial \lambda^n}\left.\langle \Psi_\lambda|\hat H(\lambda)|\Psi_\lambda\rangle \right|_{\lambda=0}
\end{align}
Using the Hellmann-Feynman theorem, this can be expressed in derivatives of the electron density $\rho$ instead\cite{von_Rudorff_2020}
\begin{align}
    E_\textrm{t}=E_\textrm{r}+\Delta E_\textrm{NN} + \int_\Omega d\mathbf{r}\sum_{n=0}^\infty \frac{\Delta V}{(n+1)!} \left.\frac{\partial^n\rho(\lambda, \mathbf{r})}{\partial\lambda^n}\right|_{\lambda=0}
\end{align}
where $\Delta E_\textrm{NN}$ is the different in nuclear repulsion energy, $\Delta V$ is the difference in the external potential created by the nuclei, and $\Omega$ is the volume that the molecule occupies. 

A formally equivalent procedure rooted in conceptual Density Functional Theory\cite{Balawender2018} does not use an explicit path $\lambda$ but rather finds the alchemical transmutation energy from electron density responses, e.g. second derivatives in energies are obtained from the alchemical hardness expressed as derivative of the linear response function $\chi$:
\begin{align}
    \frac{\partial^2E}{\partial Z_I\partial Z_J} = \int d\mathbf{r}\int d \mathbf{r'} \frac{\chi(\mathbf{r}, \mathbf{r'})}{|\mathbf{r}-\mathbf{R}_I||\mathbf{r'}-\mathbf{R}_J|}
\end{align}
with the nuclear charges $Z$ and coordinates $\mathbf{R}$ of two exemplary atoms $I, J$. Besides the technical calculation, the two approaches differ in which energy change is attributed to different atoms\cite{Balawender2018}---an unobservable quantity of diagnostic value only\cite{Rudorff2019a}---but recover exactly the same electronic energy.

All formulations assume a) that the Taylor expansion converges and b) that it converges to the true target energy, neither of which are guaranteed except for special cases like the hydrogen-like atom\cite{von_Rudorff_2020}. This work opens up a way to numerically test the convergence behavior of this Taylor expansion, regardless of the method of computation of the coefficients.

\subsection{Quantum chemistry}
Throughout this work, the restricted Hartree-Fock (HF) scheme is employed as prototypical method, as it is well-understood in its behaviour in different realms of quantum chemistry. While HF alone is not of comparable accuracy with modern quantum chemistry methods, it serves as the foundation of such, e.g. as one contribution in hybrid DFT functionals or as first step of CCSD(T) and therefore has high practical relevance. Sufficient accuracy of quantum alchemy at HF level therefore is prerequisite for its accuracy at higher levels. The perturbative approach of quantum alchemy becomes most efficient when few high-quality derivatives are employed to replace millions of lower level calculations\cite{von_Rudorff_2020}, so ultimately CCSD derivatives are desirable. HF constitutes the first step in this direction and is  well-supported in most quantum chemistry codes, allowing to test quantum alchemy in many environments.

All energy evaluations in this work have been confirmed with PySCF\cite{pyscf}. This code however does not allow for arbitrary precision math, since it interfaces Python code with Fortran and C code and no arbitrary precision library supports all three programming languages and their corresponding memory layouts. Moreoever, PySCF uses integral screening, i.e. approximating small overlap integrals with zero overlap. While such screening is highly efficient\cite{Reine2011,Rudorff2017b}, it introduces artifacts in the energy evaluation: shifting a basis center like an atom by a minute distance may already cause the overlap integral to cross the threshold. This would constitute noise if finite difference derivatives are taken. To reach arbitrary precision, such screening procedures need to be avoided, even if that increases the computational cost substantially. The focus of this work is to enable science by arbitrary precision HF calculations and derivatives rather than the efficient implementation thereof.

Based on a pure python implementation of HF\cite{meli}, the logic has been adapted for arbitrary precision math, now called APHF. Both the evaluation of the electron-electron integrals and the evaluation of the finite difference stencil have been parallelized. The code interfaces the Basis Set Exchange\cite{Pritchard2019} python library to simplify testing of different basis sets. The code for arbitrary precision HF energies, density matrices, molecular orbital energies as well as derivatives of these quantities as developed in this work is freely available online together with all input and result files used for this work\cite{APHF,data_APQA}, which allows users to test for convergence of quantum alchemy for their respective systems. Since electron-electron integrals have been implemented for s and p basis functions only, this code only supports basis sets that are limited to these basis function families for all atoms in the system. Since previous work\cite{Domenichini2020} has found STO-3G minimal basis sets to have the worst estimated convergence properties, this work focuses on these as to establish a worst-case picture. Numerical evidence both in this work as in the literature indicates that large basis sets are better suited for quantum alchemy\cite{Domenichini2020}, as the series expansion converges faster.

Unless stated otherwise, all non-PySCF calculations have been converged to the norm of the density matrix difference between consecutive steps is less than $2.5^{-998}$.

\subsection{Arbitrary precision floating point numbers}
For efficiency reasons, computers typically operate on fixed-width data types for numbers. For floating point numbers, the IEEE 754 standard is commonly used, with the most relevant variants \textit{binary32} (single precision, 32 bits) and \textit{binary64} (double precision, 64 bits) offering ~7 or ~15 significant digits, respectively. For quantum chemistry, single precision is not enough, as e.g. density matrix elements span a much larger domain. Almost all calculations are performed in double precision. 

For Taylor expansions, double precision again can be insufficient\cite{von_Rudorff_2020,Domenichini2020}, at least for higher orders. This is because a) the $n!^{-1}$ term in the Taylor series quickly grows outside the available significant digits and b) because the individual derivatives cannot be evaluated with sufficient precision so their residual errors get amplified. 

In this work, restricted HF is implemented using the mpmath\cite{mpmath} library which allows for adjustable arbitrary precision. This library also implements algorithms relevant to quantum chemistry, i.e. solving the eigenvector problem or evaluating the $\Gamma$ function which allows to calculate the Boys function in Gaussian integrals. Keeping all of the logic tailored towards arbitrary precision evaluations then allows not only alchemical derivatives but rather any evaluations of energies, density matrices and molecular orbital energies with arbitrary precision.

Unless stated otherwise in the data files, all non-PySCF calculations in this work have been performed with 1000 significant digits.

\subsection{Series expansion and convergence radius}
Multiple methods have been developed for function approximation, arguably the most popular of which are Taylor expansions where an underlying function $f(\lambda)$ is approximated by a polynomial $T_f$ the coefficients of which are obtained from the $n$-th derivatives of $f$ at the point of derivative evaluation $a$:
\begin{align}
    T_f(\lambda) \equiv \sum_{n=0}^\infty \frac{1}{n!}\left.\frac{\partial^n f(\lambda)}{\partial \lambda^n}\right|_{\lambda=a} (\lambda-a)^n
\end{align}

This expansion however may not converge at all, i.e. adding more terms drastically changes the result, or it may converge but to a value that is different from the underlying function $f(\lambda)$. In the following, the term \textit{convergence} means not only convergence of the function approximation with adding higher orders but also reaching the correct value. Typically, a Taylor series is most accurate around the point of derivative evaluation $a$. The minimal distance from that point $a$ at which the function does not converge any more is called the convergence radius $r$. While this convergence radius can be any real number, integer values are most relevant to this work, as nuclear charges are only found in discrete values. If the radius in nuclear charges becomes slightly larger than the next integer value, this opens up many more compounds to be treated by quantum alchemy at once, as the total number of possible target compounds (ignoring symmetry) which then can be assessed using the same derivatives\cite{von_Rudorff_2020} is given by $(2\left\lfloor r\right\rfloor+1)^N$. Here, each of the $N$ sites that can changed along a quantum alchemy path $\lambda$ can reach molecules with nuclear charges changed at most by $r$.

Taylor series may not converge in vicinity of singularities. In this case, a Padé approximant can be a converging alternative where the function $f$ is approximated as fraction of two polynomials:
\begin{align}
    R_f(\lambda)=\left[\sum _{j=0}^{m}a_{j}(\lambda-a)^{j}\right] \left[1+\sum _{k=1}^{n}b_{k}(\lambda-a)^{k}\right]^{-1}
\end{align}
where the coefficients $a_j$ and $b_k$ are found such that the first $n+m$ derivatives of $R_f$ and $f$ match. Common implementations including the one of mpmath\cite{mpmath} (which was used in this work) obtain these coefficients from the leading $m+n-1$ Taylor series terms. For simplicity, this work chooses $n\simeq m$.

Rigorous proofs of convergence would require analytical bounds on the derivatives $\partial_\lambda f$. In lieu of analytical bounds, the numerical values of the terms themselves can be analysed, since for most systems such analytical bounds are unavailable. In this work, two criteria are used to determine whether a series converges: a) the maximum absolute difference between the target energy including all orders up to $q$ (25$\le$q<45) and the target energy at order 45 is less than chemical accuracy, i.e. 1\,kcal/mol and b) the energy of a regular fully self-consistent HF calculation and the target energy from the series expansion at order 45 differ by at most chemical accuracy. The first condition tests whether the series itself converges to one static value and the second condition tests whether the resulting energy is correct. The limitation to 45 orders has been done for computational efficiency only, as 48 core machines have been used in this work. The code in this work allows for arbitrary precision evaluations and, consequently, arbitrary order derivatives.

\subsection{Automatic differentiation}
Instead of deriving the expression of a derivative of a function $f$ manually, automatic differentiation provides an alternative route. Here an implementation of $f$ is seen as a sequence of simple atomic operations such that repeated application of the chain rule yields the derivative. This is commonly implemented by using special data types for the input values to $f$ that keep track of the operations applied to that input throughout the execution of the function $f$. In the context of quantum alchemy, the total energy constitutes such a function: it depends on the nuclear charges, the atomic positions and the basis set. Therefore, a suitable implementation of the total energy can be instrumented with an automatic differentiation library to yield the derivatives w.r.t the nuclear charges which is what quantum alchemy requires. This process is highly non-trivial for mostly technical reasons; details can be found in previous works\cite{TamayoMendoza2018}.

The main advantage of automatic differentiation is its computational efficiency for derivatives w.r.t. to many input parameters, as compared to e.g. finite differences. However, it is still limited by machine precision, so in this work---which aims at fundamental convergence questions---, automatic differentiation numbers serve as confirmation of the leading orders only. While technically possible, a code that combines automatic differentiation and arbitrary precision would be prohibitively slow for practical applications, mostly due to the arbitrary precision operations.

\subsection{Finite differences}
Besides symbolic differentiation and automatic differentiation, a commonly applied method to obtain derivatives of a function $f$ is finite differences. In this method, small displacements of input variables are related to changes in the function value:
\begin{align}
    \frac{\partial f}{\partial \lambda}=\lim_{h\rightarrow 0} \frac{f(\lambda+h)-f(\lambda)}{h}\simeq \frac{f(\lambda+h)-f(\lambda)}{h}
\end{align}
In this work, central finite differences are used, where the expression for arbitrary orders $n$ is given as
\begin{align}
    \frac{\partial^n E}{\partial\lambda^n} = \sum _{i=0}^{n}\frac{(-1)^{i}}{h^n}{\binom {n}{i}}E\left(\lambda+\left({\frac {n}{2}}-i\right)h\right)
\end{align}
where $h$ is a small finite displacement. In practice, this becomes a weighted sum of close-by function evaluations, the stencil. The stencils have been obtained using the \textit{findiff}\cite{baer} library and have been evaluated with $h=10^{-10}$ and could be evaluated with arbitrarily small $h$ without loss of accuracy. Spot tests of finite difference derivatives obtained with smaller $h$ have shown that the orders reported here do not differ in their first 20 digits if $h$ is decreased further.

\section{Results and discussion}
\begin{figure}
    \centering
    \includegraphics[width=\columnwidth]{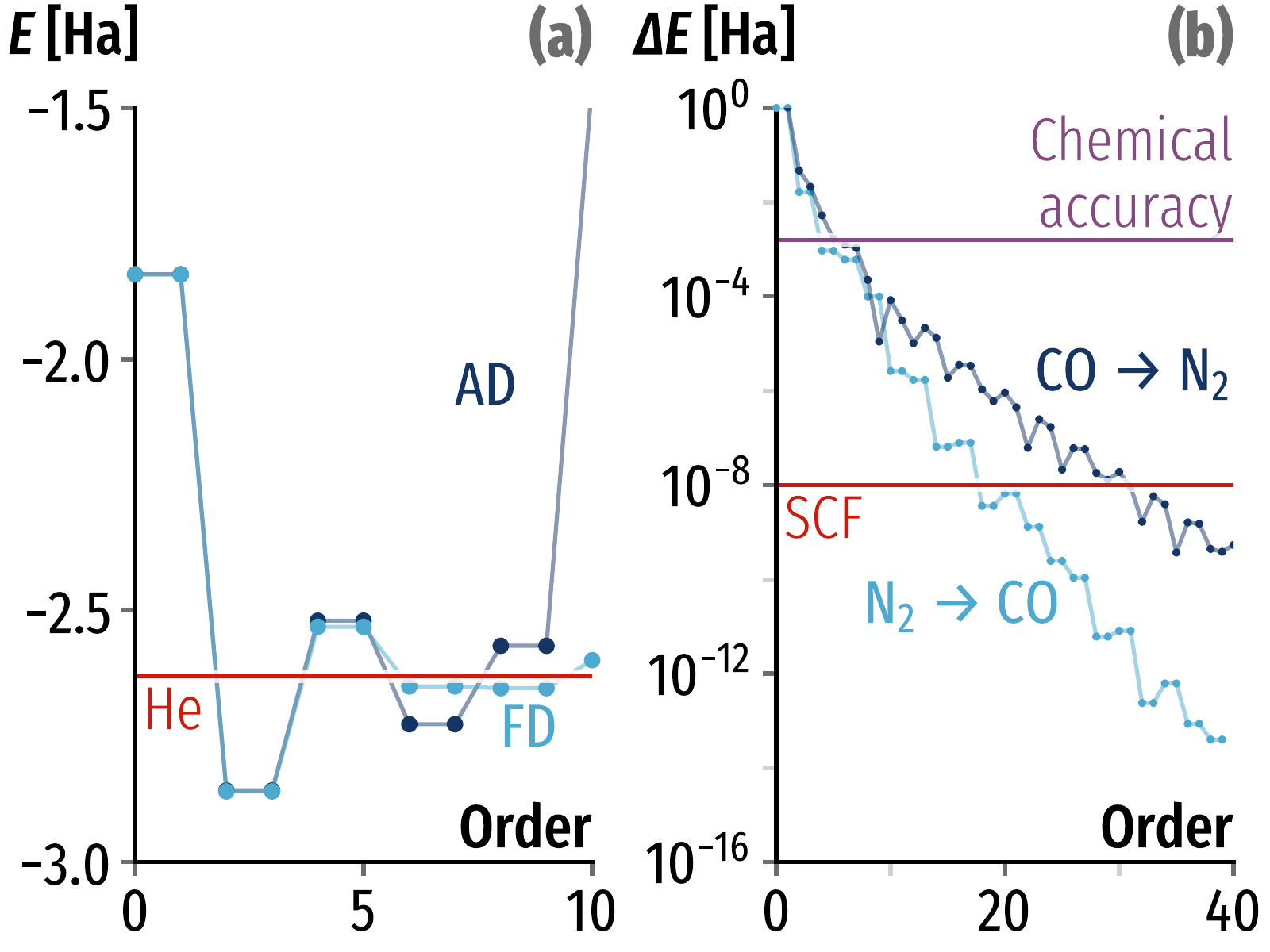}
    \caption{Convergence behaviour of a Taylor expansion in quantum alchemy with HF/STO-3G. (a) Electronic energy $E$ of He estimated from H$_2$ depending on number of perturbative orders included for the two methods automatic differentiation (AD) and finite differences (FD) where the target energy is given by a horizontal line (He). (b) The residual absolute error $\Delta E$ for estimations of CO and N$_2$ depending on number of perturbative orders included. Horizontal lines denote chemical accuracy of 1\,kcal/mol and a common SCF convergence threshold of 10$^{-8}$ Ha.}
    \label{fig:ad-fd-convergence}
\end{figure}

Figure~\ref{fig:ad-fd-convergence} shows the convergence behaviour of quantum alchemy upon inclusion of higher order perturbative terms. Panel (a) demonstrates that for the first few orders automatic differentiation and finite differences agree while for higher orders, automatic differentiation diverges which is expected due to the finite precision used in the corresponding code. Finite difference derivatives do not suffer from this problem due to the arbitrary precision code of this work and yield a Taylor series that converges towards the self-consistent target value and recovers the correct energy. This is astonishing since perturbative approaches usually are considered to be applicable if the perturbation is small, while in this case H$_2$ gets transformed into He, i.e. one site vanishes and the other site has doubled nuclear charge. 

Previous work\cite{Domenichini2020} has shown that the minimal basis set STO-3G yields less accuracy than larger basis sets for the first leading orders. Here we now see that even the minimal basis set converges to the self-consistent energy of the target system. This points towards a strong impact of the basis set on the rate of convergence and the convergence radius, which will be analysed separately in this work.

In  Figure~\ref{fig:ad-fd-convergence} (a), clear steps can be observed since all odd orders to not contribute to the final accuracy. This is due to the symmetry\cite{von_Rudorff_2020,von_Rudorff_2021} of the system and can also be observed for the case of N$_2$ $\rightarrow$ CO in panel (b) of the same figure. Since this behaviour is a consequence of the symmetry of the perturbation, this step-wise convergence is a necessary criterion for numerical stability.

Unlike previous methods, our arbitrarily precise HF code allows to evaluate much higher orders than just the leading few. Panel (b) in Figure~\ref{fig:ad-fd-convergence} shows that the residual error of quantum alchemy already after four terms can be below 1\,kcal/mol, often quoted as chemical accuracy. Furthermore, both the symmetric N$_2$ $\rightarrow$ CO case and its asymmetric inverse converges towards the self-consistent energy of the target, well beyond any SCF convergence threshold that is commonly accepted in HF calculations. While HF/STO-3G clearly is not the method of choice of modern quantum chemistry, this convergence illustrates how quantum alchemy can yield an energy of the target that is indistinguishable from the fully self-consistent energy obtained in the conventional way.

It is interesting to see that the rate of convergence in \ref{fig:ad-fd-convergence} (b) is better for the symmetric reference (N$_2$ $\rightarrow$ CO) compared to the asymmetric case. This difference gets more pronounced for higher orders. While a systematic review on the choice of suitable reference molecules is outside the scope of this work, the different rate of convergence would point towards systems of higher symmetry being the better reference system to cover a region of chemical space. It is worth noting that symmetric systems are also computationally much more efficient, since density derivatives for symmetry-equivalent atoms are identical\cite{von_Rudorff_2020}.

\begin{figure}[t]
    \centering
    \includegraphics[width=\columnwidth]{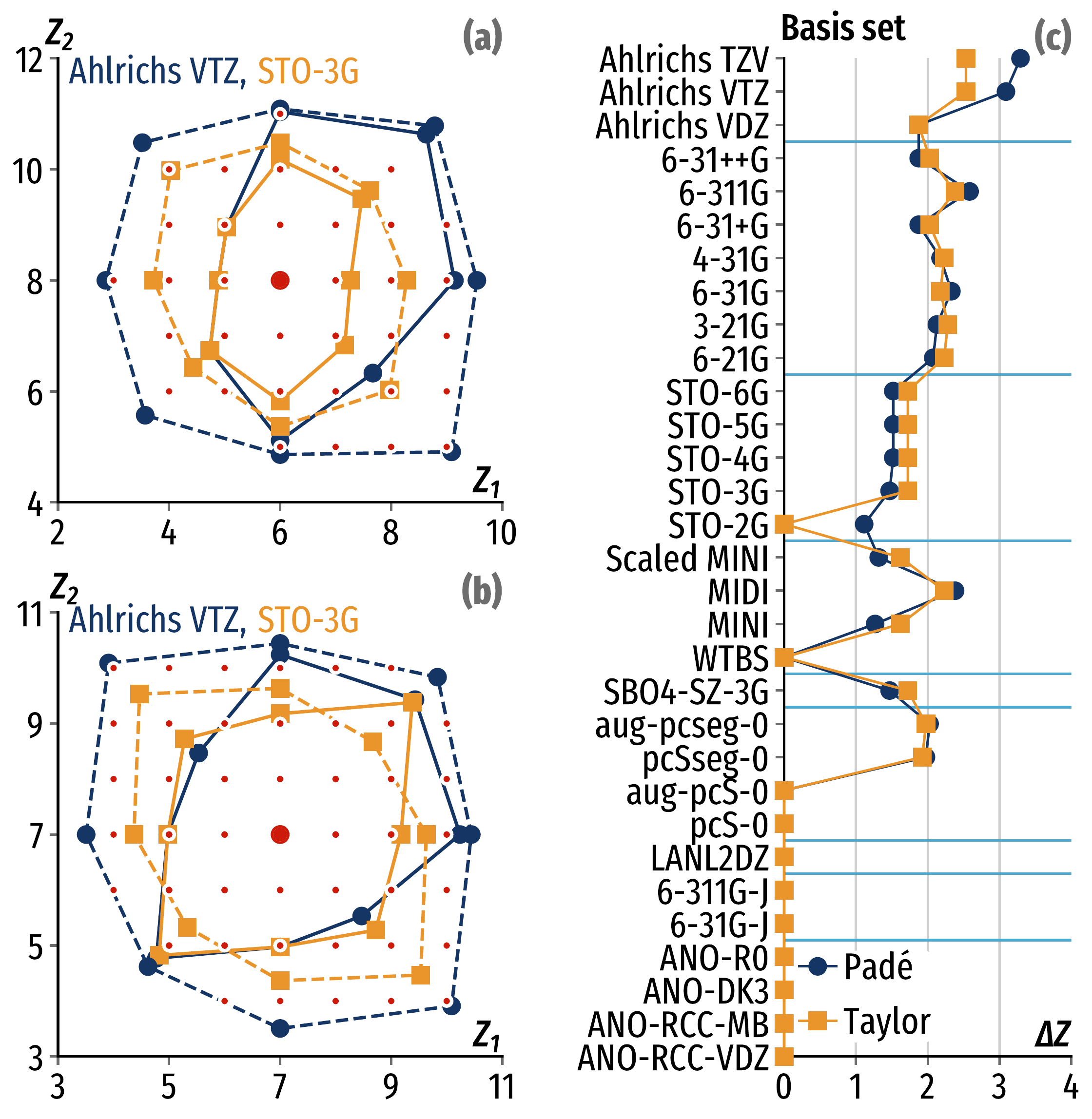}
    \caption{Convergence radii of quantum alchemy for different basis sets. (a) Using CO as reference (large red dot), eight directions of changes in nuclear charges $Z_I$ have been explicitly calculated for the Ahlrichs VTZ basis set (blue circles) and STO-3G (orange squares). Convergence radii of the Taylor series and the Padé approximant are shown as stroked and dashed lines, respectively. Integer nuclear charges corresponding to real molecules are denoted with small red dots. (b) Like (a) but for N$_2$ as reference molecule. (c) Convergence radius of N$_2$ $\rightarrow$ CO for different basis sets only containing $s$ and $p$ basis functions for both the Taylor series (squares) and the Padé approximant (circles) grouped by basis set families (Ahlrichs\cite{Schaefer1994,Schaefer1992}, Pople\cite{Binkley1980,Hehre1972,Ditchfield1971,Clark1983,Krishnan1980}, STO\cite{Hehre1969}, Huzinaga\cite{andzelm1984a,Huzinaga1990,Huzinaga1993}, Zorrilla\cite{Garcia2018}, Jensen\cite{Jensen2014,Jensen2014a,Jensen2008}, LANL\cite{dunning1977b}, Sauer\cite{Kjaer2011}, and ANO\cite{Zobel2019,Tsuchiya2001,roos2004b}), ordered by convergence radius. All numbers in all panels are evaluated with Hartree-Fock.}
    \label{fig:radius}
\end{figure}

The practical relevance of quantum alchemy is chiefly determined by the convergence radius: the more integer nuclear charges can be found within the convergence radius around the reference molecule, the more real compounds can be estimated from the perturbative expansion. As discussed in the method section even small increases of the convergence radius drastically increase the list of target compounds that can be estimated due to the combinatorial scaling with number of sites the nuclear charge of which can be changed. Figure~\ref{fig:radius} (a) shows the convergence radii in direction of different target molecules starting from CO as reference molecule. It is striking that this radius is not necessarily the same towards all target molecules, but in most cases can be extended significantly by using the Padé approximant rather than the Taylor series. The Ahlrichs VTZ basis set ((10s,6p) $\rightarrow$ [6s,3p]) has a much larger convergence radius than the minimal basis set STO-3G ((6s,3p) $\rightarrow$ [2s,1p]). Since the cost of the reference calculation in quantum alchemy is amortised by the number of targets that are estimated at once, a slightly more expensive reference calculation will often be justified to cover more target molecules at a better level of theory.

For the perturbations around the symmetric case in Figure~\ref{fig:radius} (b), we find a very similar picture, although now the convergence radii are slightly larger, in line with the observations in Figure~\ref{fig:ad-fd-convergence} (b). From a physical perspective, the figure needs to be symmetric w.r.t. the main diagonal, since the assignment of $Z_I$ is arbitrary in this symmetric case. While no such symmetry has been enforced in the calculation of the visualised data, the symmetric picture speaks to the numerical stability of the approach. It is most peculiar that the convergence radius for STO-3G along N$_2$ $\rightarrow$ O$_2^{2+}$ is much larger than the radius along N$_2$ $\rightarrow$ CO. For the Ahlrichs VTZ basis set, this is not the case. Since this is the direction that preserves symmetry in the external potential, the most likely cause for this behaviour is the inadequacy of the symmetric minimal basis set to describe strongly asymmetric electron density derivatives, in line with earlier observations\cite{Domenichini2020}.

For one direction, N$_2$ $\rightarrow$ CO, the convergence radius has been assessed systematically in Figure~\ref{fig:radius} (c). Grouped by basis set family, all basis sets from the Basis Set Exchange\cite{Feller1996,Schuchardt2007,Pritchard2019} with only $s$ and $p$ basis functions have been evaluated. Only basis sets for which HF self-consistency could be achieved within one day are shown. Interestingly, the convergence radius of quantum alchemy depends immensely on the basis set. In general, having more than one p type basis function increases the convergence radius typically by one (e.g. MIDI, 6-31G, Ahlrichs family). The lack of p type basis functions cannot be compensated by more primitives: both ANO-R0 ((13s,8p) $\rightarrow$ [2s,1p]) and WTBS ((20s, 13p) $\rightarrow$ [2,1]) are not suited for quantum alchemy. The Ahlrichs family performs very well in comparison (e.g. VDZ with [3s, 2p] has a larger convergence radius than 6-311G with [4s, 3p]), probably\cite{Domenichini2020} due to it being parametrized to have variationally optimal coefficients and exponents for free atom calculations with HF.

Considering all panels of Figure~\ref{fig:radius}, the Padé approximant typically is on par with if not better than the Taylor series for all considered basis sets. Since a larger convergence radius yields more compounds that can be estimated, future quantum alchemy applications might favor this method, even though the evaluation of a Padé approximant is slightly more expensive than the one of a Taylor series.

\begin{figure*}
    \centering
    \includegraphics[width=\textwidth]{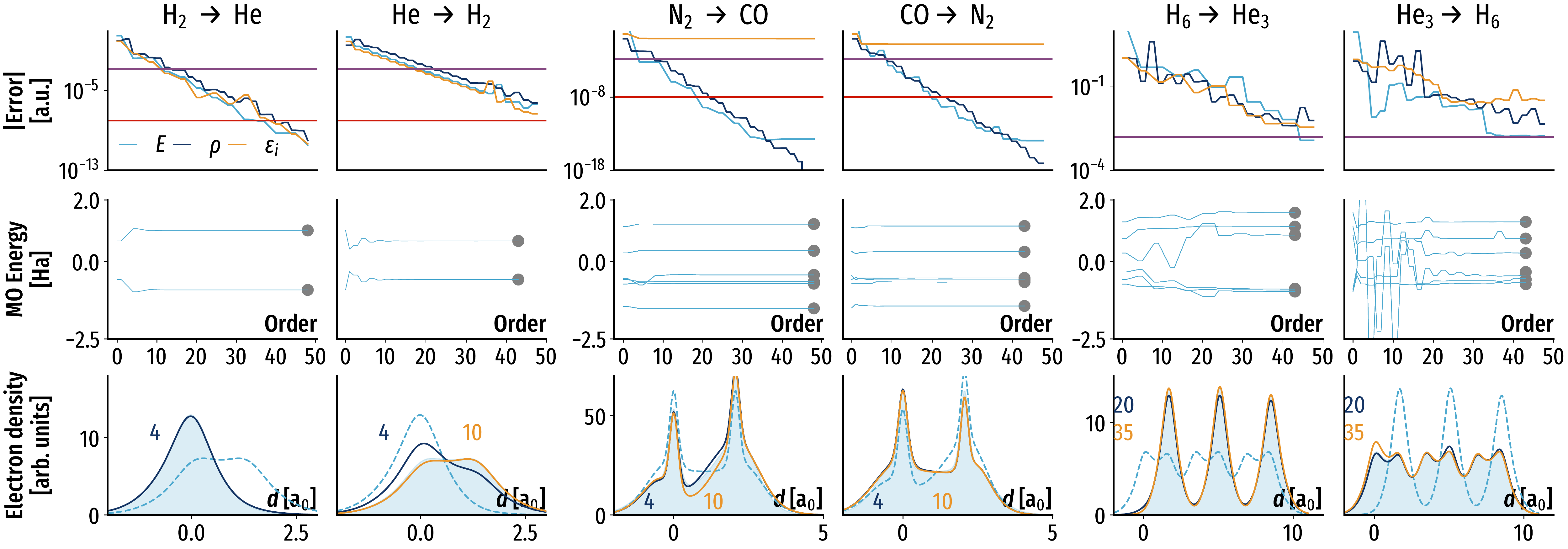}
    \caption{Accuracy of the quantum alchemy expansion depending on number of perturbation orders for electronic energies $E$, density matrix elements $\rho$ and molecular orbital energies $\epsilon_i$ for six cases (columns: from/to Hydrogen dimer, from/to Nitrogen dimer, from/to Hydrogen chain). The top row shows the absolute accuracy averaged over all density matrix elements and molecular orbitals. Horizontal lines denote chemical accuracy and SCF convergence thresholds as in Figure 1 (b). The middle row shows the individual molecular orbital energies together with their respective target value (grey dots). The bottom row shows relevant electron densities along the symmetry axis of the system: the reference (blue dashed), the target (blue filled), and densities including some given perturbation orders (lines). All data in all panels obtained with HF/STO-3G and Padé approximants.}
    \label{fig:demo}
\end{figure*}

In previous work, the focus in the evaluation of quantum alchemy has been the accuracy of the (electronic) energy. However, the Taylor expansion can be done on density matrix elements\cite{von_Rudorff_2020} and molecular orbital eigenvalues\cite{Lilienfeld2009} as well. Figure~\ref{fig:demo} shows the corresponding convergence behaviour for three pairs of reference and target molecules. The first case, H$_2$ $\leftrightarrow$ He exemplifies the large change if an atomic site is removed or created. The second case N$_2$ $\leftrightarrow$ CO illustrates the more practically relevant perturbations where the change in nuclear charge is small compared to the nuclear charges of the atoms. This would be the typical case for molecular derivatives or crystal doping. Finally, the last case H$_6$ $\leftrightarrow$ He$_3$---a dramatic change of electronic structure between reference and target---illustrates the limit of perturbations that are converging.

For the first case, the Hydrogen dimer, the symmetric reference molecule converges faster than the asymmetric case for all quantities considered. Molecular orbital energies, density matrix elements and electronic energy converge with constant rate. This is also observed for the Nitrogen dimer. Should that be a general feature of quantum alchemy for moderate practically relevant changes in nuclear charges, then this might enable accelerated convergence schemes extrapolating from a few initial expansion orders. For H$_2$ $\rightarrow$ He, both the electron density and the molecular orbital energies are practically converged after ten orders even in this case of a slowly converging\cite{Domenichini2020} minimal basis set. 

In the second example, both the density matrix coefficients and the electronic energy converge with constant rate, but not the molecular orbital energies which seem to remain at a nearly constant accuracy regardless of expansion order. This comes from the fact that in Figure~\ref{fig:demo}, all absolute molecular orbital energies accuracies are averaged. Since the two lowest lying molecular orbital energies are exchanged between reference and target, this dominates the residual average accuracy. As can be seen in the middle row in Figure~\ref{fig:demo}, the individual molecular orbital energies around the HOMO---those of greatest practical relevance---do converge to their respective target value. Since the electron density converges, this means that any one-electron property such as the electrostatic potential converges as well.

The last case tests the limits of quantum alchemy: it is well-documented that in a Hydrogen chain of an even number of equidistant atoms, dimerisation will occur\cite{Motta2020}. This is visible in the electron density profile where the electron density between sites $I$ and $I+1$ is higher if $I$ is odd. Quantum alchemy recovers this even in with a minimal basis set but convergence is very slow. Slow convergence is a sign of being close to the convergence radius, i.e. the limit of perturbations from that one reference system that are feasible. However, this extreme case of changing the electronic structure through the change of the external potential supports the assumption that for less extreme cases the Taylor series indeed converges to the correct target value. The more pronounced perturbation H$_6$ $\leftrightarrow$ Li$_2$ (which would be interesting due to the non-nuclear attractors\cite{Cioslowski1990,Terrabuio2016} forming in this case) does not converge with STO-3G.

\begin{figure}
    \centering
    \includegraphics[width=\columnwidth]{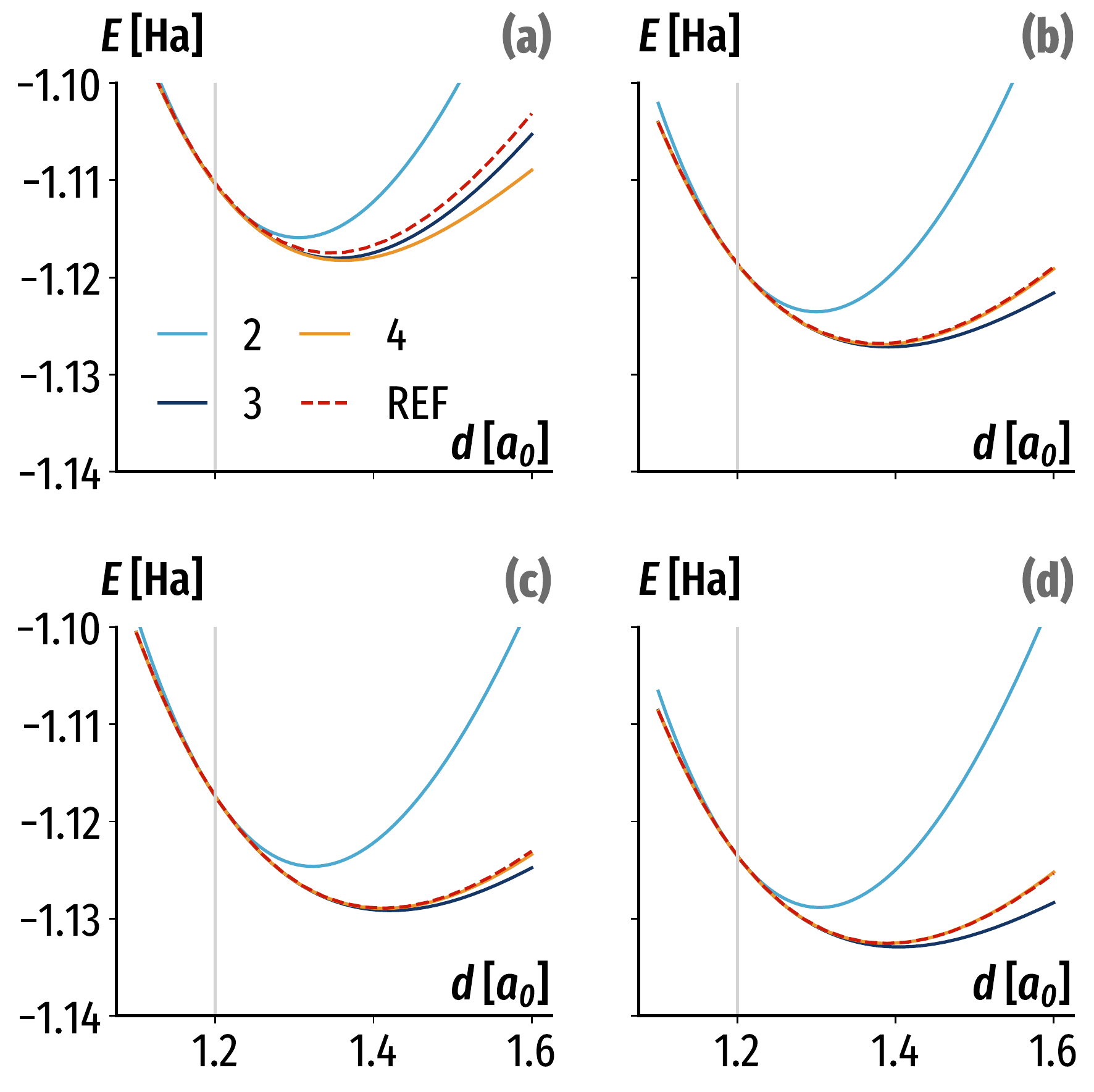}
    \caption{Potential energy surface $E$ for the H$_2$ bond stretching of distance $d$ as obtained via quantum alchemy from He using different basis sets: STO-3G (a), 6-31G (b), def2-SVP (c), def2-TZVP (d). Surfaces shown when including the first 2, 3, and 4 spatial derivatives (stroked lines) evaluated at 1.2\,$a_0$ (vertical grey line) compared to the self-consistent reference (dashed line). }
    \label{fig:spatial}
\end{figure}
In the derivation of quantum alchemy, vertical changes, i.e. unchanged geometries are assumed. This surely is a substantial restriction, since the molecular geometry will respond to the change in nuclear charges, so this relaxation is not captured yet. One way to include geometric relaxation is to calculated mixed derivatives combining the otherwise orthogonal dimensions geometry and nuclear charge. 

Considering the H$_2$ dissociation in Figure~\ref{fig:spatial}, the potential energy surface $E$ depending on the bond distance $d$ can commonly be expanded around some distance $d_0$ (here 1.2\,$a_0$). Since the first order derivative $\partial_d E$ only contains gradient information without a global minimum, the second order derivative $\partial^2_d E$ is required to formally allow for a minimum. Now these spatial derivatives in arbitrary orders can be obtained from a set of self-consistent calculations around $d_0$, using finite differences again. Each of these calculations however, can in turn be replaced by the alchemical prediction of the energy at that geometry from a He reference. In other words: Calculations on He alone can be used to predict the relaxation of H$_2$ at arbitrary orders. Keeping the number of electrons fixed, this is a generalization of the idea behind alchemical normal modes\cite{Fias2018}. Figure~\ref{fig:spatial} presents numerical evidence that this expansion converges and indeed can recover e.g. the contributions from the basis set centers which move with the bond distance $d$. For all basis sets, the third derivative already recovers the bond dissociation curve qualitatively, even though the reference distance $d_0$ around which the expansion is performed is about 15\% shorter than the equilibrium distance. At fourth order, 6-31G, def2-SVP, and def2-TZVP bond dissociation energies are indistinguishable by eye: the corresponding minima differ from the self-consistent geometry optimisation result by 0.02, 0.002, 0.002, and 0.0003 $a_0$ for STO-3G, 6-31G, def2-SVP, and def2-TZVP, respectively.

The numerical evidence for such an extreme case as the relaxation of a molecule upon separating a nucleus into two sites outlined here supports the path for calculating relaxation in quantum alchemy by alchemically predicting the spatial derivatives and following their gradient. At the same time, it appears unlikely that second order estimates are sufficient, since their potential deviates qualitatively, see Figure~\ref{fig:spatial}. If the reference structure is already closer to the relaxed minimum after alchemical perturbation than in this case or the alchemical change is less dramatic, the relevance of the third and higher order terms might decrease, effectively allowing a single Newton step to estimate the relaxed geometry. Unlike previous work\cite{Domenichini2020}, this can be done without scanning the bond distances e.g. on a regular grid and then predicting the energy from quantum alchemy, but rather by perturbing the spatial Hessian elements directly, which is much more efficient, in particular if the nuclear charges of many sites are to be changed at once. 

\section{Conclusion}
Quantum alchemy offers a systematic and rigorous way to assess many compounds in chemical spaces of same nuclear charge and same geometry. It is founded in assumptions on the nature of the electronic energy as a function of nuclear charge: convergence of the Taylor expansion and convergence thereof to the true self-consistent energy. In this work, numerical evidence for the validity of these assumptions has been obtained from an implementation of arbitrarily precise HF which in turn allowed the calculation of arbitrarily precise quantum alchemy derivatives and predictions. While the arbitrary precision math is too slow for evaluating alchemical derivatives this way for design studies, it offers a way to obtain accurate reference values and is the only available way for numerical convergence tests, since a formal proof of convergence is not available.

For the first time, alchemical convergence radii could be estimated which are key to design materials design studies based on quantum alchemy: they determine how many reference calculations are needed to cover a certain chemical space. To reach large convergence radii larger than 3 for N$_2$, suitable basis sets and the use of the Padé approximant rather than the Taylor series expansion are required. 

Even for extreme cases, the density matrix coefficients have shown to be convergent, showing that all one-electron properties which would be calculated from the density matrix alone converge as well. For several cases, the predicted energy has been shown to be more accurate than a common self-consistent energy threshold, for N$_2$, even quantum alchemy predictions more accurate than machine precision could be achieved. One such extreme case, the geometry relaxation of H$_2$ purely based on derivatives for He, exemplifies methods towards relaxing geometries with quantum alchemy.

Besides the alchemical derivatives, a code for arbitrarily precise Hartree-Fock is presented, allowing for  derivatives of energy and density matrix elements w.r.t. basis set, the basis function centers, alchemical and spatial derivatives, and all combinations thereof to arbitrary orders. This could be helpful in method development, as any derivative w.r.t any input of a Hartree-Fock calculation except the number of electrons can be evaluated this way. For example, the code provided in this work might support the development of basis sets which are more efficient for quantum alchemy, thus enabling more efficient assessment of large parts of chemical space.

\begin{acknowledgments}
This work has been supported by the von Lilienfeld lab at University of Vienna. The computational results presented have in part been achieved using the Vienna Scientific Cluster (VSC). The author thanks O. Anatole von Lilienfeld, Enrico Tapavicza, Dirk Bakowies, Max Schwilk, Simon Krug, Giorgio Domenichini, and Michael Sahre for helpful discussions.
\end{acknowledgments}

\bibliography{main}

\end{document}